\documentclass[conference]{IEEEtran}
\IEEEoverridecommandlockouts
\usepackage{cite}
\usepackage{amsmath,amssymb,amsfonts}
\usepackage{algorithmic}
\usepackage{graphicx}
\usepackage{textcomp}
\usepackage{xcolor}
\def\BibTeX{{\rm B\kern-.05em{\sc i\kern-.025em b}\kern-.08em
    T\kern-.1667em\lower.7ex\hbox{E}\kern-.125emX}}
\begin{document}

\title{Enhancing Secrecy in UAV RSMA Networks: Deep Unfolding Meets Deep Reinforcement Learning\\
}

\author{\IEEEauthorblockN{ABUZAR B. M. ADAM}
\IEEEauthorblockA{\textit{School of Communications and Information Engineering} \\
\textit{Chongqing University of Posts and Telecommunications}\\
Chongqing, China \\
email: abuzar@cqupt.edu.cn}
\and
\IEEEauthorblockN{Mohammed A. M. Elhassan}
\IEEEauthorblockA{\textit{School of Computer Science} \\
\textit{Zhejiang Normal University}\\
Jinhua, Zhejiang, China \\
email:Mohammedac29@zjnu.edu.cn}

}

\maketitle

\IEEEpubidadjcol

\begin{abstract}
In this paper, we consider the maximization of the secrecy rate in multiple unmanned aerial vehicles (UAV) rate-splitting multiple access (RSMA) network. A joint beamforming, rate allocation, and UAV trajectory optimization problem is formulated which is nonconvex. Hence, the problem is transformed into a Markov decision problem and a novel multiagent deep reinforcement learning (DRL) framework is designed. The proposed framework (named DUN-DRL) combines deep unfolding to design beamforming and rate allocation, data-driven to design the UAV trajectory, and deep deterministic policy gradient (DDPG) for the learning procedure. The proposed DUN-DRL have shown great performance and outperformed other DRL-based methods in the literature.
\end{abstract}

\begin{IEEEkeywords}
Secrecy rate maximization, rate-splitting multiple access, deep reinforcement learning, deep unfolding.
\end{IEEEkeywords}

\section{Introduction}
Recently, rate-splitting multiple access (RSMA) has received great attention as a promising non-orthogonal technology for the sixth generation (6G) networks \cite{b1}. RSMA is a powerful interference management and multiple access technology that improves the system performance by dividing the transmitted message into public message that can be decoded by all users and private message that can only be decoded by the intended user \cite{b2}.

To improve the coverage and connectivity in densely-deployed networks, unmanned aerial vehicle (UAV) has been considered in many research works \cite{b3}. UAV combined with multiple antennas can improve the energy efficiency of the network \cite{b4}. However, due to the broadcast nature of the UAV communication channels, they are vulnerable to malicious attacks \cite{b5}. Therefore, designing efficient and secure networks is crucial issue for next generation networks.

Machine learning methods have been recently used to improve the network performance in terms of security \cite{b6} and resource allocation \cite{b7,b8,rev1,rev2}. Deep reinforcement learning (DRL) is one of machine learning techniques used in UAV-aided networks \cite{b9,b10}. DRL frameworks have shown great performance in solving several problems. However, most of deep neural networks used in the literature are mainly data-driven \cite{b11}. Aiming at incorporating the recent advances in deep unfolding \cite{bb40} with DRL and motivated by the above discussion on RSMA and secrecy issues, we design a DRL framework-named DUN-DRL-to handle the secrecy rate maximization in multiple UAV-aided RSMA network. We can summarize our contributions as follows
\begin{itemize}
  \item We consider the secrecy rate maximization in multiuser multiple UAV RSMA network. We formulate an optimization problem with secrecy and UAV trajectory constraints. We decouple the problem into rate allocation and beamforming subproblem and UAV trajectory subproblem.
  \item We transform the subproblems into Markov decision problem and we design multiagent deep deterministic policy gradient (DDPG)-based DRL framework. To build the beamforming and rate allocation agent, we design an iterative solution and convert it into multi-layer neural network via deep unfolding while UAV trajectory agent is a convolutional neural network (CNN).
  \item The performance of the proposed DUN-DRL is verified via simulation results. We show that the proposed DUN-DRL outperforms other well-known DRL frameworks in the literature.
\end{itemize}

\section{System Model and Problem Formulation}
We consider a downlink RSMA where $U$ UAVs are acting as aerial base stations (ABSs) and serving $K$ legitimate users in presence of $I$ eavesdroppers. Let ${\cal U} = \left\{ {1,...,U} \right\},{\cal K} = \left\{ {1,...,K} \right\},$ and ${\cal I} = \left\{ {1,...,I} \right\}$ represent the set of ABSs, legitimate users, and eavesdroppers, respectively. All ABSs are equipped with $M$ antennas, while legitimate users and the eavesdroppers are equipped with $N$ antennas. ${{\bf{L}}_k} = {\left( {{x_k},{y_k}} \right)^T},{{\bf{L}}_i} = {\left( {{x_i},{y_i}} \right)^T}$ are respectively the locations of the legitimate user $k$ and the eavesdropper $i$. The flight period $\rm T$ of the UAV is divided into $T$ time slots with $\delta_t$. Hence, the location of the ABS $u$ at the time slot $t$ is given as ${\bf{q}}\left( t \right) = {\left( {{x_u}\left( t \right),{y_u}\left( t \right),{H_u}\left( t \right)} \right)^T}$.

ABS $u$ employs RSMA and sends a common signal ${s_c}\left( t \right) \in {\mathbb{C}^{d \times 1}}$ with $d$  represents the number data streams to the its associated users with normalized power which is precoded with the beamformer ${{\bf{w}}_c}\left( t \right) \in {\mathbb{C}^{N \times d}}$ and private message ${s_k}\left( t \right) \in {\mathbb{C}^{d \times 1}}$ to its associated user $k$ which is mapped with the beamformer ${{\bf{w}}_k}\left( t \right) \in {\mathbb{C}^{N \times d}}$. The received signal at the user $k$ served by the ABS $u$ is given as
\begin{equation}\label{eqn1}
\begin{aligned}
{y_k}\left[ t \right] &= \underbrace {{\bf{h}}_{u,k}^H\left( t \right){{\bf{w}}_c}\left( t \right){s_c}\left( t \right)}_{\text{Common signal}}\\
 &+ \underbrace {{\bf{h}}_{u,k}^H\left( t \right){{\bf{w}}_{u,k}}\left( t \right){s_k}\left( t \right)}_{\text{Desired private signal}}\\
 &+ \underbrace {\sum\nolimits_{v \ne u} {\sum\nolimits_{j = 1,j \ne k}^K {{\bf{h}}_{v,k}^H\left( t \right){{\bf{w}}_{v,j}}\left( t \right){s_j}\left( t \right)} } }_{\text{Interference}} + \underbrace {{{\bf{z}}_{u,k}}}_{\text{AWGN}},
\end{aligned}
\end{equation}
where ${\bf{h}}_{u,k}^H\left( t \right) \in {^{N \times M}}$ is the channel between ABS $u$ and user $k$ and  ${{\bf z}_{u,k}} \sim {\cal C}{\cal N}\left( {0,\sigma _{u,k}^2{{\bf{I}}_d}} \right)$ is additive while Gaussian noise (AWGN). The signal-to-interference-plus-noise ratio (SINR) to decode the common message is given by
\begin{equation}\label{eqn2}
{\gamma _c}\left( t \right) = \frac{{\left\| {{\bf{h}}_{u,k}^H\left( t \right){{\bf{w}}_c}\left( t \right)} \right\|_F^2}}{{{{\bf{R}}_{c,u,k}}\left( t \right) + \sum\nolimits_{k = 1}^K {\left\| {{\bf{h}}_{u,k}^H\left( t \right){{\bf{w}}_{u,k}}\left( t \right)} \right\|_F^2} }},
\end{equation}
where ${{\bf{R}}_{c,u,k}}\left( t \right) = \sigma _{u,k}^2{{\bf{I}}_d} + \sum\nolimits_{v \ne u} {\sum\nolimits_{j = 1,j \ne k}^K {\left\| {{\bf{h}}_{v,k}^H\left( t \right){{\bf{w}}_{v,j}}\left( t \right)} \right\|_F^2} } $. Hence, the rate for decoding the common message is given as
\begin{equation}\label{eqn3}
{R_c}\left( t \right) = {\log _2}\det \left( {{\bf{I}} + {\gamma _c}\left( t \right)} \right),
\end{equation}
All the legitimate users should be capable of decoding the common message. To guarantee that, the constraint $\sum\nolimits_{k = 1}^K {{\mathfrak{r}_{u,k}}\left( t \right)}  \le \mathop {\min }\limits_k {R_c}\left( t \right)$ is imposed. Similarly, the rate of user $k$ to decode its private signal is given as
\begin{equation}\label{eqn4}
{R_{u,k}}\left( t \right) = {\log _2}\det \left( {{\bf{I}} + {\gamma _{u,k}}\left( t \right)} \right),
\end{equation}

where
\begin{equation}\label{eqn5}
{\gamma _{u,k}}\left( t \right) = \frac{{\left\| {{\bf{h}}_{u,k}^H\left( t \right){{\bf{w}}_{u,k}}\left( t \right)} \right\|_F^2}}{{{{\bf{R}}_{u,k}}\left( t \right) + \sum\nolimits_{l = 1,l \ne k}^K {\left\| {{\bf{h}}_{u,k}^H\left( t \right){{\bf{w}}_{u,l}}\left( t \right)} \right\|_F^2} }},
\end{equation}
and
\begin{equation}\label{eqn6}
{{\bf{R}}_{u,k}}\left( t \right) = \sigma _{u,k}^2{{\bf{I}}_d} + \sum\nolimits_{v \ne u} {\sum\nolimits_{j = 1,j \ne k}^K {\left\| {{\bf{h}}_{v,k}^H\left( t \right){{\bf{w}}_{v,j}}\left( t \right)} \right\|_F^2} } ,
\end{equation}
The rate of the eavesdropper $i$ who is trying to wiretap the communication link between ABS $u$ and user $k$ is given as
\begin{equation}\label{eqn7}
{R_{u,i}}\left( t \right) = {\log _2}\det \left( {{\bf{I}} + {\gamma _{u,i}}\left( t \right)} \right)
\end{equation}
where
\begin{equation}\label{eqn8}
{\gamma _{u,i}}\left( t \right) = \frac{{\left\| {{\bf{h}}_{u,i}^H\left( t \right){{\bf{w}}_{u,k}}\left( t \right)} \right\|_F^2}}{{{{\bf{R}}_{u,i}}\left( t \right) + \sum\nolimits_{l = 1,l \ne k}^K {\left\| {{\bf{h}}_{u,i}^H\left( t \right){{\bf{w}}_{u,l}}\left( t \right)} \right\|_F^2} }},
\end{equation}
and
\begin{equation}\label{eqn9}
{{\rm{R}}_{u,i}}\left( t \right) = \sigma _{u,i}^2{{\bf{I}}_d} + \sum\nolimits_{v \ne u} {\sum\nolimits_{j = 1,j \ne k}^K {\left\| {{\bf{h}}_{v,i}^H\left( t \right){{\bf{w}}_{v,j}}\left( t \right)} \right\|_F^2} } ,
\end{equation}
The secrecy rate is then defined as follows
\begin{equation}\label{eqn10}
R_{u,k}^{\sec }\left( t \right) = \mathfrak{r}_{c,u,k}^{\sec }\left( t \right) + {\left[ {{R_{u,k}}\left( t \right) - {R_{u,i}}\left( t \right)} \right]^ + }
\end{equation}
where $\sum\nolimits_{k = 1}^K {\mathfrak{r}_{c,u,k}^{\sec }\left( t \right)}  \le {R_c}\left( t \right) - {R_{c,i}}\left( t \right)$ is a non-negative optimization variable which implies ${R_{c,i}}\left( t \right) \le {R_c}\left( t \right)$ and ${R_{c,i}}\left( t \right)$ is achievable rate of decoding the common signal message by the eavesdropper $i$ and given as
\begin{equation}\label{eqn11}
{R_{c,i}}\left( t \right) = {\log _2}\det \left( {{\bf{I}} + {\gamma _{c,i}}\left( t \right)} \right)
\end{equation}
where
\begin{equation}\label{eqn12}
{\gamma _{c,i}}\left( t \right) = \frac{{\left\| {{\bf{h}}_{u,i}^H\left( t \right){{\bf{w}}_c}\left( t \right)} \right\|_F^2}}{{{{\bf{R}}_{c,u,i}}\left( t \right) + \sum\nolimits_{k = 1}^K {\left\| {{\bf{h}}_{u,i}^H\left( t \right){{\bf{w}}_{u,k}}\left( t \right)} \right\|_F^2} }},
\end{equation}
And ${{\bf{R}}_{c,u,i}}\left( t \right) = \sigma _{u,i}^2{{\bf{I}}_d} + \sum\nolimits_{v \ne u} {\sum\nolimits_{j = 1,j \ne k}^K {\left\| {{\bf{h}}_{v,i}^H\left( t \right){{\bf{w}}_{v,j}}\left( t \right)} \right\|_F^2} }$. We aim at maximizing the secrecy rate of network by optimizing the beamformer, rate, and UAV trajectory. Thus, we formulate the following optimization problem

\begin{subequations}\label{eqn13:main}
\begin{align}
&\mathop {\max }\limits_{{\bf{w}},\mathfrak{r},{\bf{Q}}} && \sum\nolimits_{u \in {\cal U}} {\sum\nolimits_{k = 1}^K {R_{u,k}^{\sec }\left( t \right)} } &   & \tag{\ref{eqn13:main}}\\
& \text{s.t.}&& \sum\nolimits_{k = 1}^K {\mathfrak{r}_{c,u,k}^{\sec }\left( t \right)}  \le {R_c}\left( t \right) - {R_{c,i}}\left( t \right),\label{eqn13:a} \\
&            &&\left\| {{{\bf{w}}_c}\left( t \right)} \right\|_F^2 + \sum\nolimits_{k = 1}^K {\left\| {{{\bf{w}}_{u,k}}\left( t \right)} \right\|_F^2}  \le {P_u},\label{eqn13:b}\\
&            &&\mathfrak{r}_{c,u,k}^{\sec }\left( t \right) \ge 0,\label{eqn13:c}\\
&            &&{\left\| {{{\bf{q}}_u}\left( {t + 1} \right) - {{\bf{q}}_u}\left( t \right)} \right\|^2} \le {D^2},t = 1,...,T - 1,\label{eqn13:d}\\
&            &&{{\bf{q}}_u}\left( 0 \right) = {{\bf{q}}_0},t = 1,...,T - 1,\label{eqn13:e}\\
&            &&{\left\| {{{\bf{q}}_u}\left( T \right) - {\bf{q}}_u^F} \right\|^2} \le {D^2},\label{eqn13:f}
\end{align}
\end{subequations}
Constraints \eqref{eqn13:a} is the secrecy constraints, \eqref{eqn13:b} is the power budget constraint, and \eqref{eqn13:c} is the non-negativity constraint of the secrecy optimization variable. \eqref{eqn13:d}, \eqref{eqn13:e}, and \eqref{eqn13:f} are the UAV movement constraints. Problem (15) is non-convex due to the coupling of the parameters in \eqref{eqn13:a} and \eqref{eqn13:b}. The problem is also non-smooth due to the operator ${\left[  \cdot  \right]^ + }$. Even with fixed beamformer and rate allocation, optimizing the problem with respect to the UAV trajectory is highly complex. Therefore, aiming at jointly handling problem \eqref{eqn13:main}, we resort to DRL. In the next section, we design a DRL-based architecture to tackle problem \eqref{eqn13:main}.

\section{Deep Unfolding-based Deep Reinforcement Learning }
\begin{figure}[htbp]
\centerline{\includegraphics[width=3.2in]{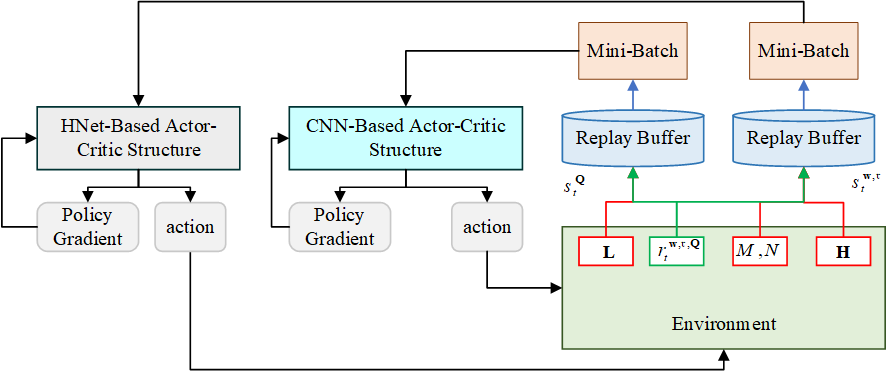}}
\caption{Structure of the proposed DUN-DRL}
\label{model}
\end{figure}

In order to design a deep unfolding-based model, we need to provide an algorithm to unfold. In our case, we design an iterative solution of the beamforming and rate allocation, then we transform that solution into a multi-layer deep neural network via deep unfolding named HNet (i.e., hybrid net). However, for UAV trajectory, we design a CNN-based actor-critic network. Both models are used to design the proposed deep unfolding-based DRL (DUN-DRL).  Fig. \ref{model}, shows the proposed architecture of the DU-DRL. In the below subsections present the steps of designing DUN-DRL.

\begin{figure}[htbp]
\centerline{\includegraphics[width=3.2in]{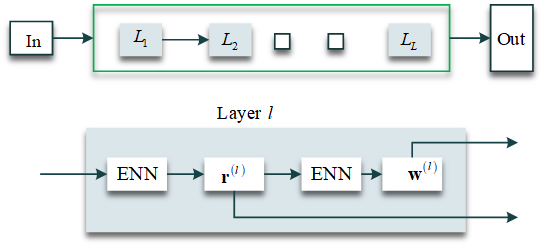}}
\caption{Detailed structure of HNet}
\label{hnet}
\end{figure}

\subsection{Design of Beamforming and Rate Allocation Network (HNet) via Deep Unfolding}
The beamforming and rate allocation subproblem is given as follows
\begin{subequations}\label{eqn14:main}
\begin{align}
&\mathop {\max }\limits_{{\bf{w}},\mathfrak{r}} && \sum\nolimits_{u \in {\cal U}} {\sum\nolimits_{k = 1}^K {R_{u,k}^{\sec }\left( t \right)} } &   & \tag{\ref{eqn14:main}}\\
& \text{s.t.}&& \eqref{eqn13:a}, \eqref{eqn13:b},\eqref{eqn13:c},\notag
\end{align}
\end{subequations}
The problem is non-convex due to the coupling. We define the auxiliary $\zeta  = {\left[ {{\zeta _{1,1}},{\zeta _{1,2}},...,{\zeta _{U,K}}} \right]^T}$ and $\upsilon  = {\left[ {{\upsilon _{1,1}},...,{\upsilon _{U,I}}} \right]^T}$ and we remove for simplicity, the problem is then reformulated as
\begin{subequations}\label{eqn15:main}
\begin{align}
&\mathop {\max }\limits_{{\bf{w}},\mathfrak{r},\zeta ,\upsilon } && \sum\nolimits_{u \in {\cal U}} {\sum\nolimits_{k = 1}^K {\mathfrak{r}_{c,u,k}^{\sec } + {{\left[ \begin{array}{l}
{\log _2}\left( {1 + {\zeta _{u,k}}} \right)\\
 - {\log _2}\left( {1 + {\zeta _{u,i}}} \right)
\end{array} \right]}^ + }} } &   & \tag{\ref{eqn15:main}}\\
& \text{s.t.}&&\eqref{eqn13:b}, \eqref{eqn13:c},\notag\\
&            &&\sum\nolimits_{k = 1}^K {\mathfrak{r}_{c,u,k}^{\sec }}  \le {\log _2}\left( {1 + {\upsilon _c}} \right) - {\log _2}\left( {1 + {\upsilon _{c,i}}} \right),\label{eqn15:a}\\
&            &&{\upsilon _c} \le \frac{{\left\| {{\bf{h}}_{u,k}^H{{\bf{w}}_c}} \right\|_F^2}}{{{{\bf{R}}_{c,u,k}} + \sum\nolimits_{k = 1}^K {\left\| {{\bf{h}}_{u,k}^H{{\bf{w}}_{u,k}}} \right\|_F^2} }},\label{eqn15:b}\\
&            &&{\upsilon _{c,i}} \le \frac{{\left\| {{\bf{h}}_{u,i}^H{{\bf{w}}_c}} \right\|_F^2}}{{{{\bf{R}}_{c,u,i}} + \sum\nolimits_{k = 1}^K {\left\| {{\bf{h}}_{u,i}^H{{\bf{w}}_{u,k}}} \right\|_F^2} }},\label{eqn15:c}\\
&            &&{\zeta _{u,k}} \le \frac{{\left\| {{\bf{h}}_{u,k}^H{{\bf{w}}_{u,k}}} \right\|_F^2}}{{{{\bf{R}}_{u,k}} + \sum\nolimits_{l = 1,l \ne k}^K {\left\| {{\bf{h}}_{u,k}^H{{\bf{w}}_{u,l}}} \right\|_F^2} }},\label{eqn15:d}
\end{align}
\end{subequations}
Using Theorem 2 in \cite{b12}, the optimal solution of \eqref{eqn15:main} should obey the following
\begin{equation}\label{eqn16}
\sum\nolimits_{k = 1}^K {\mathfrak{r}_{c,u,k}^{\sec }}  - {\log _2}\left( {1 + {\upsilon _c}} \right) + {\log _2}\left( {1 + {\upsilon _{c,i}}} \right) = 0
\end{equation}
\begin{equation}\label{eqn17}
\resizebox{0.97\hsize}{!}{${\rm Tr}\left( {\underbrace {{\upsilon _c}\left( {{{\bf{R}}_{c,u,k}} + \sum\nolimits_{k = 1}^K {{\bf{h}}_{u,k}^H{{\bf{w}}_{u,k}}{\bf{w}}_{u,k}^H{{\bf{h}}_{u,k}}} } \right)}_{{{\bf{M}}_1}} - \underbrace {{\bf{h}}_{u,k}^H{{\bf{w}}_c}{\bf{w}}_c^H{{\bf{h}}_{u,k}}}_{{{\bf{M}}_2}}} \right) = 0,$}
\end{equation}
\begin{equation}\label{eqn18}
\resizebox{0.97\hsize}{!}{${\rm Tr}\left( {\underbrace {{\upsilon _{c,i}}\left( {{{\bf{R}}_{c,u,i}} + \sum\nolimits_{k = 1}^K {{\bf{h}}_{u,i}^H{{\bf{w}}_{u,k}}{\bf{w}}_{u,k}^H{{\bf{h}}_{u,i}}} } \right)}_{{{\bf{M}}_3}} - \underbrace {{\bf{h}}_{u,i}^H{{\bf{w}}_c}{\bf{w}}_c^H{{\bf{h}}_{u,i}}}_{{{\bf{M}}_4}}} \right) = 0,$}
\end{equation}
\begin{equation}\label{eqn19}
\resizebox{0.97\hsize}{!}{${\rm Tr}\left( {\underbrace {{\zeta _{u,k}}\left( {{{\bf{R}}_{u,k}} + \sum\nolimits_{l = 1,l \ne k}^K {{\bf{h}}_{u,k}^H{{\bf{w}}_{u,l}}{\bf{w}}_{u,l}^H{{\bf{h}}_{u,k}}} } \right)}_{{{\bf{M}}_5}} - \underbrace {{\bf{h}}_{u,k}^H{{\bf{w}}_{u,k}}{\bf{w}}_{u,k}^H{{\bf{h}}_{u,k}}}_{{{\bf{M}}_6}}} \right) = 0,$}
\end{equation}
With fixed  $\mathfrak{r},\upsilon ,\zeta$, the beamforming subproblem can be expressed as
\begin{subequations}\label{eqn20:main}
\begin{align}
&\text{find} && {\bf w} &   & \tag{\ref{eqn20:main}}\\
& \text{s.t.}&&\eqref{eqn13:b}, \eqref{eqn16},\notag\\
&            &&{{\bf{M}}_1} - {{\bf{M}}_2} = {{\bf{0}}_{N \times d}}, \label{eqn20:a}\\
&            &&{{\bf{M}}_3} - {{\bf{M}}_4} = {{\bf{0}}_{N \times d}}, \label{eqn20:b}\\
&            &&{{\bf{M}}_5} - {{\bf{M}}_6} = {{\bf{0}}_{N \times d}}, \label{eqn20:c}\\
&            &&{\rm rank}\left( {\bf{w}} \right) = d, \label{eqn20:d}
\end{align}
\end{subequations}
Applying Cholesky decomposition on ${{\bf{M}}_1}$, ${{\bf{M}}_3}$ and ${{\bf{M}}_5}$, we have the following closed-form solutions in \eqref{eqn21} and \eqref{eqn22} on top of next page.
\begin{figure*}
\begin{equation}\label{eqn21}
{{\bf{w}}_c} = \frac{{{\bf{C}}_1^{ - 1}{v_{1:d}}\left( {{{\left( {{\bf{C}}_1^H} \right)}^{ - 1}}{{\bf{M}}_2}{\bf{C}}_1^{ - 1}} \right)}}{{{{\left\| {{\bf{C}}_1^{ - 1}{v_{1:d}}\left( {{{\left( {{\bf{C}}_1^H} \right)}^{ - 1}}{{\bf{M}}_2}{\bf{C}}_1^{ - 1}} \right)} \right\|}_F}}} - \frac{{{\bf{C}}_2^{ - 1}{v_{1:d}}\left( {{{\left( {{\bf{C}}_2^H} \right)}^{ - 1}}{{\bf{M}}_4}{\bf{C}}_2^{ - 1}} \right)}}{{{{\left\| {{\bf{C}}_2^{ - 1}{v_{1:d}}\left( {{{\left( {{\bf{C}}_2^H} \right)}^{ - 1}}{{\bf{M}}_4}{\bf{C}}_2^{ - 1}} \right)} \right\|}_F}}},
\end{equation}
\begin{equation}\label{eqn22}
{{\bf{w}}_{u,k}} = \frac{{{\bf{C}}_3^{ - 1}{v_{1:d}}\left( {{{\left( {{\bf{C}}_3^H} \right)}^{ - 1}}{{\bf{M}}_6}{\bf{C}}_3^{ - 1}} \right)}}{{{{\left\| {{\bf{C}}_3^{ - 1}{v_{1:d}}\left( {{{\left( {{\bf{C}}_3^H} \right)}^{ - 1}}{{\bf{M}}_6}{\bf{C}}_3^{ - 1}} \right)} \right\|}_F}}},
\end{equation}
\end{figure*}
where ${\bf{C}}_1$, ${\bf{C}}_2$ and ${\bf{C}}_3$ are triangular matrices from Cholesky decomposition of ${{\bf{M}}_1}, {{\bf{M}}_3}$, and ${{\bf{M}}_5}$, respectively. ${v_{1:d}}\left(  \cdot  \right)$ is used to extract the first dominant eigenvector of the given matrix. Assuming strong duality, constraints \eqref{eqn15:b}, \eqref{eqn15:c}, and \eqref{eqn15:d}. Hence, we have
\begin{equation}\label{eqn23}
\sum\nolimits_{k = 1}^K {\mathfrak{r}_{c,u,k}^{\sec }}  = {\log _2}\left( {1 + {\upsilon _c}} \right) - {\log _2}\left( {1 + {\upsilon _{c,i}}} \right),
\end{equation}
Using the solutions in in \eqref{eqn21}, \eqref{eqn22}, and \eqref{eqn23}, we design the model in Fig. \ref{hnet}. ENN is a fully connected neural network used to perform the extraction of the first dominant vector and it is built as in \cite{bb50}.

\subsection{CNN-based Network for UAV trajectory}
The proposed CNN-based network for UAV trajectory consists of three convolutional layers with $3\times3$ convolutional kernels and each layer is followed by a batch normalization and rectified linear unit (ReLU) function. Then, a max pooling layer followed by linear layer for the output.

In order to maximize the secrecy rate using the proposed architecture in Fig. \ref{model}, Markov decision is used reformulate the problem by defining the state space, actions, and rewards for each single agent at each time slot $t$ as explained below

For the beamforming and rate allocation, the state is $s_t^{{\bf{w}},r}$ at the time slot $t$ includes the channel state information of legitimate users and the eavesdroppers  ${\bf{H}} = \left\{ {\left\{ {{{\bf{h}}_{u,k}}} \right\}_{k = 1}^K,\left\{ {{{\bf{h}}_{u,i}}} \right\}_{i = 1}^I} \right\}$ at the time slot $t$ and the number of antennas $M, N$. The action $a_t^{{\bf{w}},r}$ represents both the beamforming and rate allocation. Then, the reward  $r_t^{{\bf{w}},\mathfrak{r},{\bf{Q}}}$ which is the same for the UAV trajectory as well is given as follows \cite{b13}.
\begin{equation}\label{eqn23}
r_t^{{\bf{w}},\mathfrak{r},{\bf{Q}}} = \tanh \left( \begin{array}{l}
\sum\nolimits_{u \in {\cal U}} {\sum\nolimits_{k = 1}^K {R_{u,k}^{\sec }} }  - {C_1}{\rho _\mathfrak{r}} - {C_2}{\rho _{\bf{w}}}\\
 - {C_3}{\rho _{ -\mathfrak{r} }} - {C_4}{\rho _{{{\bf{q}}_1}}} - {C_5}{\rho _{{{\bf{q}}_0}}} - {C_6}{\rho _{{{\bf{q}}_F}}}
\end{array} \right),
\end{equation}
where ${\rho _\mathfrak{r}}, {\rho _{\bf{w}}}, {\rho _{ -\mathfrak{r} }}, {\rho _{{{\bf{q}}_1}}}, {\rho _{{{\bf{q}}_0}}},$ and ${\rho _{{{\bf{q}}_F}}}$ are penalties associated with the constraints.
For the case of the UAV trajectory, the state $s_t^{\bf{Q}}$ includes the locations of users and UAVs ${\bf{L}} = \left\{ {\left\{ {{{\bf{L}}_k}} \right\}_{k = 1}^K,\left\{ {{{\bf{L}}_i}} \right\}_{i = 1}^I,\left\{ {{{\bf{q}}_u}} \right\}_{u = 1}^U} \right\}$ at each time slot $t$. The action in this case is the UAV movement ${{\bf{q}}_u}\left( {t + 1} \right) - {{\bf{q}}_u}\left( t \right) = {\mu _u}\left( t \right)\theta \left( t \right)$ where ${\mu _u}\left( t \right)$ and $\theta \left( t \right)$ respectively denote the flying distance and the direction. Finally, the reward is same as \eqref{eqn23}.

\section{Numerical Results}\label{AA}
To simulate the results, we considered two UAVs each is serving 4 legitimate users in presence of 2 eavesdroppers trying to wiretap the communication links between the legitimate users and the UAVs. $P_u = 3$ dBm, $\delta_t$ = 0.1, $T = 1$ s, $\sigma_{u,k} = -80$ dBm, $M = 4$, $N = 2$, $d = 2$, pathloss exponent is 3.5, carrier frequency is 28 GHz, antenna spacing  ${\lambda  \mathord{\left/
 {\vphantom {\lambda  2}} \right.
 \kern-\nulldelimiterspace} 2}$, UAV velocity is 20 m/s.

For agents' architectures, the number of HNet blocks is 6, and ENN has 3 hidden layers. Batch size is 128, and learning rate is 0.001. The parameters above are default unless stated otherwise.

For the comparison, first, we considered the global optimal solution in \cite{b14} for the beamforming and rate allocation in addition to exhaustive search for UAV trajectory to form globally optimal solution for the problem. Second, we considered TDDRL method in \cite{b13} which is similar to our method. Finally, we have a single agent DRL method.

\begin{figure}[htbp]
\centerline{\includegraphics[width=3.1in]{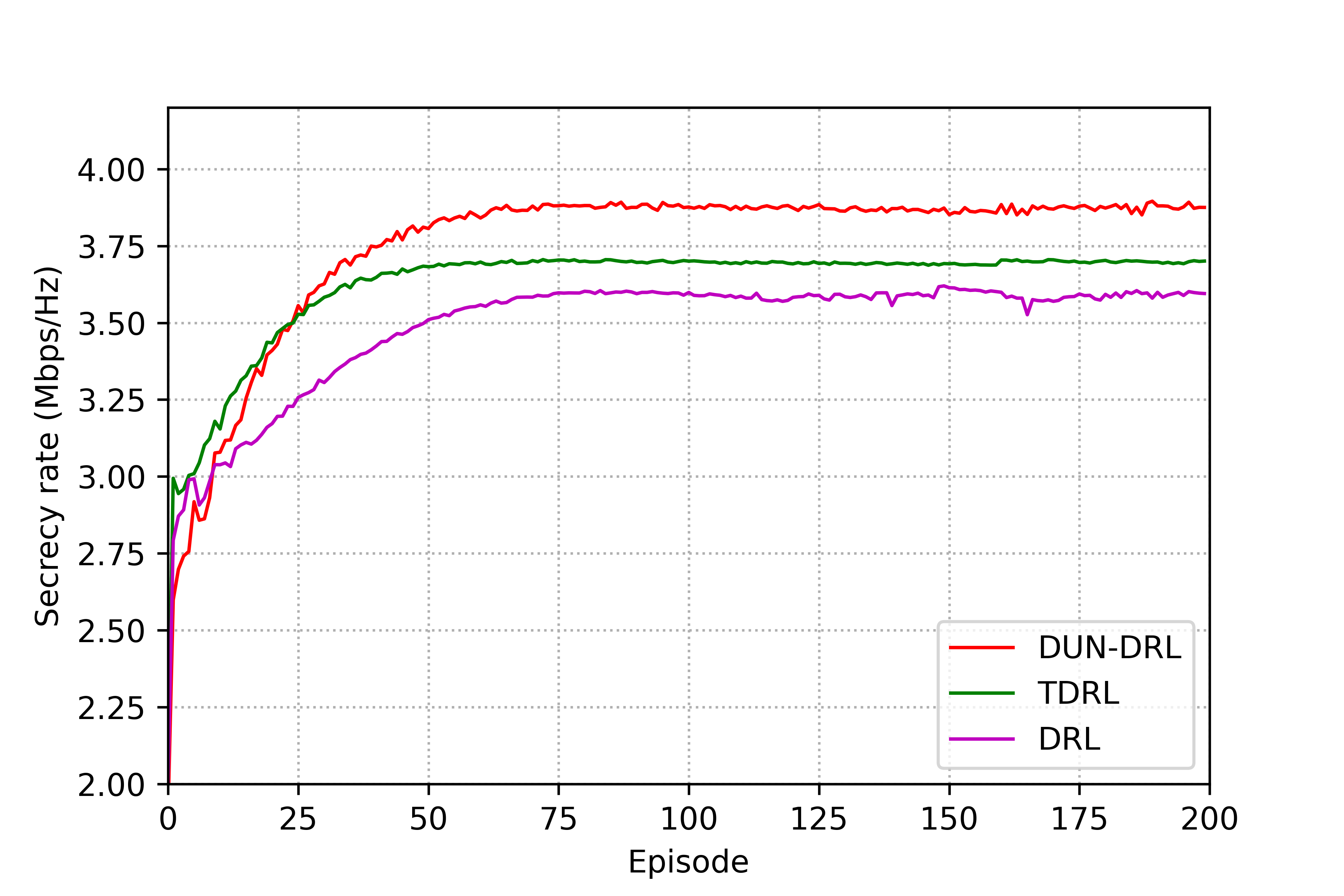}}
\caption{Achievable secrecy rate versus training episodes.}
\label{fig1}
\end{figure}

\begin{figure}[htbp]
\centerline{\includegraphics[width=3.1in]{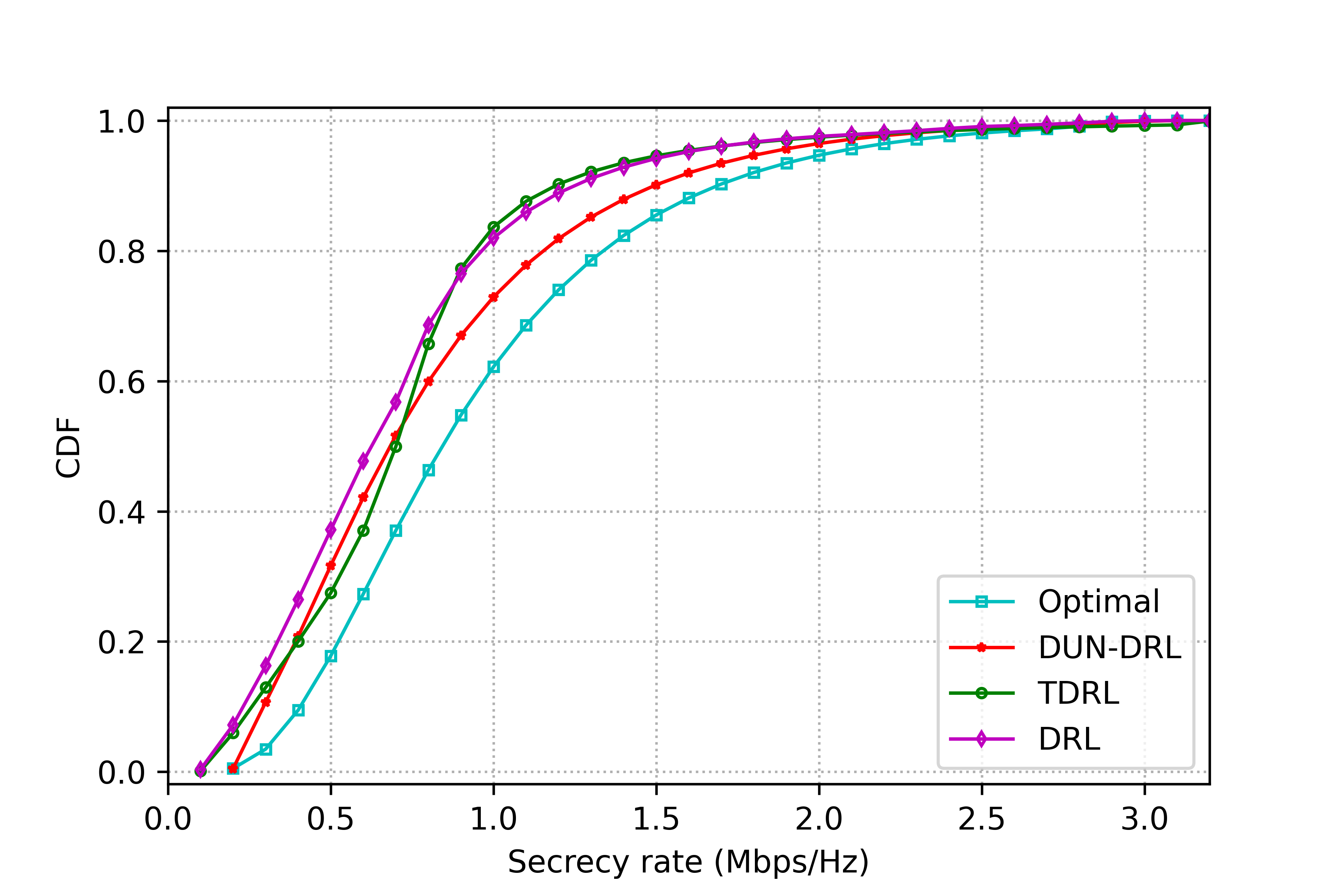}}
\caption{Cumulative distribution functions (CDF) of the secrecy rate.}
\label{fig2}
\end{figure}
Fig. \ref{fig1} shows the achievable secrecy rate versus the number of training episodes. The advantageous structure of the proposed DUN-DRL provides better performance since it combines both data-driven and the working mechanism of iterative solution. From Fig. \ref{fig1} it is obvious that both DUN-DRL and TDDRL converge in fewer number of episodes and achieve higher secrecy rate compared to DRL model and that is because of the severe and continuous changes in the channel state information as well as the locations $\bf L$. However, DUN-DRL achieves between performance and tends to satisfy the constraints better than TDDRL. This is very obvious from Fig. \ref{fig2} where we can observe that DUN-DRL tends to achieve higher rates for users and the number of users with lower rates is smaller compared with TDDRL and DRL, and closer to that of the optimal solution.

\begin{figure}[htbp]
\centerline{\includegraphics[width=3.1in]{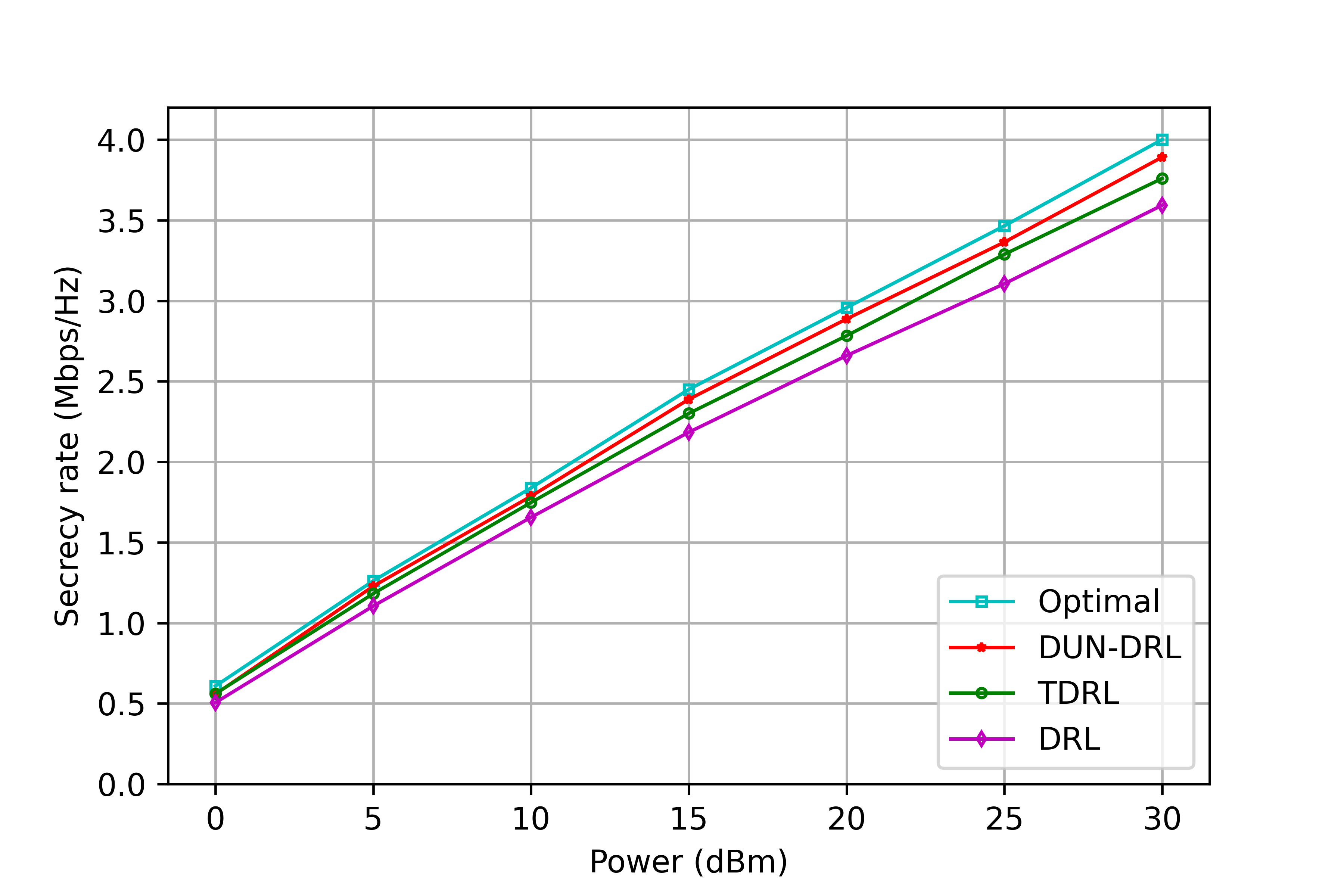}}
\caption{Achievable secrecy rate versus UAV power.}
\label{fig5}
\end{figure}

To achieve these higher secrecy rates, the power consumption in case of DUN-DRL is less compared to that of TDDRL and DRL. This is demonstrated in Fig. \ref{fig5} where DUN-DRL achieves 98.06\% performance compared with the optimal solution. While TDDRL and DRL achieve 95.81\% and 92.55\%, respectively.
\section{Conclusion}
In this work, we investigated the secrecy rate maximization in multiple UAV RSMA networks. We formulate the optimization problem which is nonconvex. Since it is challenging to jointly optimize the beamforming, rate allocation, and UAV trajectory, we transform the problem into a Markov decision problem. Then, a novel multiagent DRL framework named DUN-DRL is designed by combining deep unfolding for beamforming and rate allocation, CNN-based framework for UAV trajectory optimization while DDPG mechanism is adopted as a learning mechanism. The simulation results showed that the proposed DUN-DRL outperforms other frameworks in the literature.


\end{document}